\documentstyle[12pt]{article}

\newcommand{\rf}[1]{(\ref{#1})}
\newcommand{\beq}{\begin{equation}}
\newcommand{\eeq}{\end{equation}}
\newcommand{\bea}{\begin{eqnarray}}
\newcommand{\eea}{\end{eqnarray}}

\newcommand{\n}{\nu}
\newcommand{\m}{\mu}
\newcommand{\th}{\theta}

\newcommand{\vf}{\varphi}

\begin{document}

\addtolength{\baselineskip}{0.20\baselineskip}
\hfill    NBI-HE-93-17

\begin{center}

\vspace{36pt}

{\large{\bf Electroweak Magnetism, W-condensation
and Anti-Screening}}\footnote{Based on an invited
talk at the 4th Hellenic  School on
Elementary Particle Physics, Corfu September 92.
To appear in the proceedings.}

\vspace{36pt}

{\sl J. Ambj\o rn and P. Olesen} \\

\vspace{12pt}

 The Niels Bohr Institute\\
Blegdamsvej 17, DK-2100 Copenhagen \O ,\\
 Denmark

\end{center}

\vfill

\begin{center}
{\bf abstract}
\end{center}

\vspace{12pt}

\noindent
We review how external magnetic fields act as perfect
probes of the non-abelian nature of the electroweak theory.

\vfill

\newpage

{\Large{\bf 1. Introduction}}
\vspace{.5cm}

It has been difficult to get an intuitive feeling of
asymptotic freedom and the related concept of anti-screening
which is so essential for modern physics. It is our intention
to show that the ingredients for a semiclassical understanding
of these concepts are present precisely in the electroweak theory.
One of the major problems in achieving a semiclassical understanding of
antiscreening is that any semiclassical picture is related to
the  low-energy region of the theory, and  this sector
is very complicated in a theory like QCD, precisely because the
asymptotic freedom and anti-screening make the interaction strong and
the vacuum structure complicated.
It turns out that  external magnetic fields in the electroweak theory
act as  perfect monitors of non-abelian excitations. Contrary to
QCD the vacuum of the electroweak theory is trivial.
As an external magnetic field is turned on, there will be
a threshold above which the external magnetic field will interact
with the non-abelian $SU(2)_w$ weak fields. This is due to the fact that
the Weinberg angle $\th_{w}$ is different from zero and electromagnetism
consequently is a mixture of the abelian hypercharge sector $U(1)_y$ and
the non-abelian sector $SU(2)_w$.  If the energy in the magnetic field
is comparable with the one needed to restore the $SU(2)_w \times U(1)_y$
symmetry of the electroweak theory we will gradually see the
non-abelian aspects of the theory and we can analyse the excitations
in detail.

In the following we will give a non-technical description of electroweak
magnetism, with special emphasis on the formation of a $W$-condensate
and its anti-screening properties.

\vspace{.7cm}

{\Large{\bf 2. The instability}}
\vspace{.5cm}

QED in principle allows the existence of arbitrary large magnetic
fields. This is to be contrasted with the case of electric fields
where external electric fields are able to perform work on the
virtual electrons and put them on-shell. For a magnetic field we
know that the Lorentz force cannot perform any work on charged
particles, and the ``vacuum'' consisting of an external magnetic
field is indeed a stable field configuration. The quantum fluctuations
of charged scalar and spinor particles will in general try to screen the
external field and the resultant effective action for the external
field will be non-linear but there will never be a pair production
of real particles as is the case if we have strong electric fields.

The situation is different in the electroweak theory.  Let us
describe why: Consider first a charged scalar particle moving
in an constant, external magnetic field. Classically the particle can move
freely in the direction of the magnetic field while  it will perform
harmonic oscillations in the plane perpendicular to the external field.
In the quantum field theory this leads to the following energy levels:
\beq
E^2_n(scalar) = k^2_3 + (2n + 1)eH + m^2 \label{1.9}
\eeq
\noindent
where the numbers $n$ label the famous Landau levels.

The only difference for the spinor particle is the coupling between
the spin and the magnetic field:
\beq
E^2_n(spinor) = k^2_3 + (2n + 1)eH - 2eH \cdot s + m^2 \label{1.10}
\eeq
where $s$ can take the values $\pm \frac{1}{2}$.
The factor 2 present in front of the coupling $H \cdot s$-coupling
reflects that the $g$-factor of the electron is two.

It is seen that the quantum fluctuations are stable since $E^2_n > 0$
both for scalar and spinor particles.
However, it should be noticed that coupling of the external field
and the spin can lower the energy relative to that of a scalar
particle.
{}From \rf{1.10} it is seen that this gain in energy is exactly equal to
the kinetic zero point energy of the harmonic oscillator in the
plane perpendicular to the external field
and in the massless case one has a zero mode.
\footnote{This zero mode is in fact the origin of the anomaly in
QED, see for instance \cite{2}.} when the Landau level $n=0$.

In the electroweak theory we have spin one particles, the $W$'s and the
$Z$'s. The $W$-particles are charged and will interact with an external
magnetic field, much like a spinor particle. In fact the energy levels
in the constant external field will be given by:
\beq
E^2_n(vector) = k^2_3 + (2n+1) eH - 2eH \cdot s + m^2. \label{1.13}
\eeq
The only difference compared to the spinor case
is that the spin $s$ takes the values
$-1,0,+1$.
{\it However, this implies that large magnetic fields are unstable
since $E^2$ can be less than zero if $H > H^{(1)}_{crit}$, where}
\beq
eH^{(1)}_{crit} = m^2 \label{1.14}
\eeq

\vspace{12pt}
\noindent
In the case of the spinor particle the $g$-factor 2 follows from the
Dirac equation. In fact one of the main achievements of the Dirac
equation is precisely to explain why the $g$-factor of the electron
is 2. For a spin one particle there is no obvious reason why the
$g$-factor should be 2. We get a $g$ factor 1 if we couple a charged
vector particle minimally to electrodynamics. Let us now shown that the
$g$-factor 2 is a direct consequence of the non-abelian nature of
$SU(2)_w$.

The Lagrangian of the electroweak theory is (ignoring fermions):

\bea
{\cal L} = & - & \{\frac{1}{2} |\tilde{D}_{\mu}W_{\nu} - \tilde{D}_{\nu}
                     W_{\mu}|^2 + \frac{1}{4} f^2_{\mu \nu} +
                     \frac{1}{4} Z^2_{\mu \nu} + (\partial_{\mu}\varphi)^2\}
                      \nonumber \\
           & - & \{\frac{g^2 \varphi^2}{2} W^{\dagger}_{\mu}W_{\mu}
               + \frac{1}{2} \frac{g^2\varphi^2}{\cos^2 \theta}
                \frac{1}{2} Z^2_{\mu} - 2\lambda\varphi^2_0\varphi^2\}
                 \nonumber \\
           & - & ig(f_{\mu\nu} \sin \theta +
                      Z_{\m\n} \cos \theta)\; W^{\dagger}_{\mu}
                 W_{\nu} \nonumber \\
           & - & {\frac{1}{2} g^2 \left((W^{\dagger}_{\mu}W_{\mu})^2
                 - W^2_{\mu}W^{\dagger^2}_{\mu}\right)
                 + \lambda(\varphi^4 + \varphi^4_0)} \label{2.1}
\eea

\noindent
where $W_{\mu}$ and $Z_{\mu}$ are the usual vector boson fields and
the covariant derivative is given by:
\beq
\tilde{D}_{\mu} = \partial_{\mu} - ig(A_{\mu} \sin \theta +
          + Z_{\mu} \cos \theta) \label{2.2}
\eeq

The electromagnetic vector potential is denoted $A_{\mu}$ and the field
strength corresponding to electromagnetic currents and neutral currents
are
\beq
f_{\mu\nu} = \partial_{\mu}A_{\nu} - \partial_{\nu}A_{\mu}~~~~,
       ~~~~Z_{\mu\nu} = \partial_{\mu}Z_\nu - \partial_{\nu}Z_{\mu}
             \label{2.3}
\eeq
\noindent
The electromagnetic charge $e$ and the hypercharge $g'$ are related to
the $SU(2)_w$-charge $g$ by the standard relations
\beq
e = g \sin\theta_w~~~~,~~~~g' = g \tan \theta_w \label{2.3a}
\eeq

The Lagrangian \rf{2.1} is written in the {\it unitary} gauge
where the Higgs field is real. If the Higgs field has no zeroes
this is clearly possible since we can always by an $SU(2)_w$ gauge
transformation $U_{\varphi}$ make sure that
\beq
U_{\phi}
\left( \begin{array}{c} \phi_1 \\ \phi_2 \end{array}  \right)
   =  \left(\begin{array}{c} 0 \\ \phi_R \end{array} \right) \label{2.3b}
\eeq
\beq
\phi_R = \sqrt{|\phi_1|^2 + |\phi_2|^2} \label{2.3c}
\eeq
The various terms in \rf{2.1} have been grouped as kinetic terms,
mass terms, magnetic moment terms and fourth-order terms.

The coupling of the $W_{\mu}$ field to an external electromagnetic
field $A^{ex}_{\mu}$ will then be given by
\beq
{\cal L}(W) = -\frac{1}{4} (f_{\mu\nu}^{ex})^2
               - \frac{1}{2} |D_{\mu}W_{\nu} - D_{\nu}W_{\mu}|^2
              - m^2_w W^{\dagger}_{\mu}W_{\mu}
              - ief^{ex}_{\mu\nu}W^{\dagger}_{\mu}W_{\nu} \label{2.5}
\eeq
\noindent
where
\bea
D_{\mu} & = & \partial_{\mu} - ieA^{ex}_{\mu} \nonumber \\
      e & = & g \sin \theta \nonumber \\
    m^2_w & = & \frac{g^2\varphi^2_0}{2} \label{2.6}
\eea
The important term in \rf{2.5} is the ``anomalous'' magnetic moment term
$ief^{ex}_{\mu\nu} W^{\dagger}_{\mu}W_{\nu}$, compared to a minimal
coupled theory of charged vector particles. This term changes precisely
the $g$-factor of the $W$-particle from one to two, and
the origin of this term can be traced back to the non-Abelian nature
of $SU(2)_{w}$, as is clear from \rf{2.1}.
Amusingly, it was noticed already in the forties (see \cite{4})
that such an ``anomalous'' term is natural when one couples a charged
vector particle to the electromagnetic field.

\vspace{.7cm}

{\Large{\bf 3. The electroweak transition at $H^{(1)}_{crit}$}}
\vspace{.5cm}

It is clear that a constant magnetic field is a solution
to the classical equations of motion.
We simply choose $W_{\mu} = Z_{\mu} = 0,~~\varphi = \varphi_0$.
We saw in the last section that $H^{ex}$ cannot be arbitrary large.
Eventually the fluctuations of the charged field $W_{\mu}$ will be
so large an instability will appear and
the system undergoes a phase transition to a new
phase where there is a $W_{\mu}$ (and $Z_{\mu}$) condensate.

The origin and the general form of the unstable mode of \rf{1.13} can
be understood from the following considerations.
The instability of the linearized theory \rf{2.5} for large magnetic
fields is due to the anomalous magnetic moment term.
We assume that the electromagnetic field is
in the $\hat{z}$-direction in space, but the arguments
to be given are clearly valid for any field configurations if only
the spatial variation is slow compared to the field strength.
The assumption implies that $f_{12}~~(= H)$ is the only field
component different from zero and we get an effective mass term
\beq
(W^{\dagger}_{1},W^{\dagger}_{2})
\left(\begin{array}{ll}
m^{2}_{w}  & ieH \\
-ieH       & m^{2}_{w}
\end{array} \right)
\left(\begin{array}{c} W_1 \\ W_2 \end{array} \right)
\eeq \label{2.10}
The mass eigenvalues are
\beq
m^2 = m^2_w \pm eH \label{2.11}
\eeq
and it is seen that the lowest mass becomes tachyonic above a critical
field strength
\beq
eH^{(1)}_c = m^2_w \label{2.12}
\eeq
\noindent
and the corresponding eigenvector is
\beq
(W_1,W_2) = (W,iW) \label{2.13}
\eeq
\noindent
satisfying that the kinetic term in \rf{2.5} is equal to zero if
\beq
D_iW_j - D_jW_i = 0 \label{2.14}
\eeq
\noindent
or
\beq
(D_1 + iD_2)W(x_1,x_2) = 0 \label{2.15}
\eeq
\noindent
It is easily checked that \rf{2.15} is indeed the general solution
corresponding to the lowest eigenvalue $E_n$ of \rf{1.13}.
The general solution to \rf{2.15} in the case of a constant
magnetic field is
\bea
W(x_1,x_2) & =  & e^{- \frac{1}{2}eH x^2_1}F(z) \label{2.16} \\
z & = & x_1 + ix_2 \label{2.17}
\eea
\noindent
where $F$ is an arbitrary analytic function.
The degeneracy of the eigenvalues $E_n$ is due to the translational
invariance perpendicular to $H$.

{}From these considerations it is natural to expect that for
external fields larger than or equal to $H^{(1)}_c$ one shall
have a $W$ (and $Z$) condensate with $W_2$ and $W_1$ related by a
constant phase $\pi/2$.
If we choose $F(z) = e^{\frac{1}{4}z^2m^2_w}$ we get a
``vortex'' configuration
\beq
|W(x_1,x_2)| = e^{-\frac{1}{4} m^2_w(x^2_1 + x^2_2)} \label{2.18}
\eeq
where the $W$-fields are localized around $x_1 = x_2 = 0$ and
circles around the magnetic line at $x_1 = x_2 = 0$.

Since $W$ is assumed small (we ignore $W^4$ terms in
\rf{2.1}) we are considering a linearized approximation and
one can (and should) make superposition
of such vortex solutions to form a real condensate in order to
counteract the instability all over space.
If a condensate of vortices is formed as in a type II superconductor
it is natural to expect some kind of periodicity.
If we demand $|W(x_1,x_2)|$ to be periodic, it can be shown that
$F(z)$ must be a generalized Jacobi theta function with parameters
depending on the unit cell of periodicity in the $x_1-x_2$ plane
(see \cite{5} for details).
The area ${\cal A}$ of the unit cell of periodicity will be linked
to the field strength of the external magnetic field by the
relation
\beq
{\cal A} = \frac{2\pi}{eH} \label{2.19a}
\eeq
\noindent
In the case where the unit cell is a square of side length
$L$ we have \cite{5}:
\bea
F(z) & = & \vartheta (z/L)        \label{2.19b} \\
\vartheta (z) & = &\Sigma^{\infty}_{n = - \infty}
                  e^{- \pi n^2 + 2\pi n z} \label{2.20}\\
L& =& \sqrt{2\pi/eH} \label{2.20a}
\eea
\noindent

At this point it is interesting to compare the results to the
ones of a type II superconductor near the upper critical field
strength $H_c$.
The Ginzburg-Landau Lagrangian for a type II superconductor is
\beq
{\cal L}(\phi ,A) = - \frac{1}{4} f^2_{\mu\nu} -
                     |D_{\mu}\psi|^2 - \lambda(\psi^2 - \psi^2_0)^2
                     \label{2.21}
\eeq
\noindent
where $\psi$ is the complex order parameter for the superconductor.
For notational simplicity we will assume that $\psi$ carries a
charge $e$ (although in a real superconductor the Cooper pair
will carry a charge $2e$ and the relativistic version \rf{2.21} is
not so relevant).
The ground state of \rf{2.21} will be $\psi=\psi_0$ and the excitations
of the condensate will correspond to a Higgs mass
\beq
m^2_h = 2e^2\psi^2_0  \label{2.22a}
\eeq
\noindent
In an external magnetic field $H^{ex}$ there is an upper critical
field strength $H_c$ above which the symmetry is restored.
We loose superconductivity ($\psi=0)$ and the photon become
massless.
The upper critical field strength is given by
\beq
eH_c = \frac{1}{2}m^2_h   \label{2.23}
\eeq
\noindent
Near the upper critical field strength $H_c$ the amplitude of
$\psi$ is small and we can to a good approximation  consider only
quadratic terms in \rf{2.21}.
The linearized equation of motion for $\psi$ is
\beq
-D^2_{\mu} \psi - 2\lambda \psi^2_0 \psi = 0 \label{2.24}
\eeq
\noindent
It is seen that \rf{2.24} is identical to the linearized equations
for $W$ at $H_c$ since we can write
\beq
\left\{- (D_1 - iD_2)(D_1 + iD_2) - (\frac{1}{2} m^2_H - eH_c)\right\}
\psi = 0 \label{2.24a}
\eeq
\noindent
or
\beq
(D_1 + iD_2)\psi \approx 0 \label{2.24b}
\eeq
\noindent
At $H_c$ the condensate solution for $\psi$ will have the same
analytical form as $W$.
Especially, the area of a unit cell of periodicity will be given
by \rf{2.19a}.

The $W$-condensate differs from the ordinary $\psi$-condensate of a
type II superconductor in the sense that the stable solution is
$f_{12} = H,~W =0$ for low fields while the condensate is only
formed for ${f}_{12} > H^{(1)}_{crit}$.
For a superconductor.
the solution is $f_{12} = H, ~\psi=0$ ~~for~~ $f_{12} > H_{crit}$
and a $\psi$ condensate is only formed for the applied field less than
$H_{crit}$.
Near $H_{crit}$ we have in both cases that the strength of the
condensate is small and it makes sense to study the back reaction
of the condensate on the electromagnetic field.
We have from the equations of motion:
\beq
\partial_{\nu}f_{\nu\mu} = - j^{induced}_{\mu} \label{2.25}
\eeq
\noindent
where $j^{induced}_{\mu}$ for the two Lagrangians \rf{2.21} and
\rf{2.5} are given by:
\bea
j^{induced}_{\mu}(\psi) & = & - ie \psi^{\dagger}
(\stackrel{\rightarrow}{D}_{\mu}
                           - \stackrel{\leftarrow}{D}^{\dagger}_{\mu})\psi
                                \label{2.26} \eea
\bea
j^{induced}_{\mu}(W_{\mu}) & = & - ieW^{\dagger}_{\nu}
(\stackrel{\rightarrow}{D}_{\mu} -
\stackrel{\leftarrow}{D}^{\dagger}_{\nu})W_\nu  \nonumber \\
{}~&~ & +\left\{ ieW^{\dagger}_{\nu}(D_{\nu}W_{\mu}
- (D_{\nu}W_{\mu})^{\dagger} W_{\nu}
- ie \partial_{\nu}(W^{\dagger}_{\mu} W_{\nu} - W_{\mu}
W^{\dagger}_{\nu})\right\} \label{2.27}
\eea
\noindent
In \rf{2.27} we have divided the induced current into a convective part
and a spin part (see also \cite{7}).
As a consequence of $D_{\mu}W_{\mu} = 0$ and the ansatz \rf{2.13} we get
\beq
j^{induced}_{\mu} (W) = 2ieW^{\dagger}(\stackrel{\rightarrow}{D}_{\mu} -
\stackrel{\leftarrow}{D}^{\dagger}_{\mu})W   \label{2.28}
\eeq
\noindent
and we now see a crucial difference between the ordinary $\psi$-condensate
and the $W$-condensate.
It is well known that the $\psi$-condensate will set up currents
to screen the external magnetic field (Lenz law).
In the linearized approximation $W=\psi$, and since \rf{2.28} is
identical to \rf{2.26} {\it except} for a sign, we conclude the
$W$-condensate set up currents to {\it enhance} the external
magnetic field.
We have {\it anti-screening}.

We can trace the difference to the spin part of the induced current.
This part is minus two times the convective current part.
It also tells us that this anti-screening is a consequence of
asymptotic freedom.
The 1-loop vacuum polarization can be calculated from the
current-current correlation
\beq
\pi_{\mu\nu}(x-y) = <j^{induced}_{\mu}(x) j^{induced}_{\nu}(y)>
                                \label{2.29}
\eeq
\noindent
and the $\beta$-function can be related to the renormalization
$Z_3$ of $\pi_{\mu\nu}$ in the so-called background gauge.
The contributions to $\beta$-function from the convective part and from the
spin part of the current $j^{induced}_{\mu}(x)$ can be
explicitly calculated (see \cite{7}) and the negative
sign of the $\beta$-function is due to the spin-current correlation.

\vspace{.7cm}

{\Large{\bf 4. Symmetry restoration at $H^{(2)}_{crit}$}}
\vspace{.5cm}

The strength of the condensate, the magnitude of $W$, was not
determined in the linearized approximation of section 3. It
is to be expected that the growth in $W$ due to the instability
will be stabilized by including the $|W|^4$ terms for the
Lagrangian \rf{2.1}.
This is what happens in a type II superconductor.
The strength of $|\psi|^2$ is eventually determined by including
the $\lambda|\psi|^4$ term from \rf{2.21}.

Let us for simplicity still ignore the $Z_{\mu}$ field.
In the presence of an electromagnetic field $f_{\mu\nu}$ the ``potential''
energy involving the Higgs- and the $W$-field is $(W_{\mu}
= (W,iW,0,0)$
\bea
V(\varphi,W) & = & 2(ef_{12}-m^2_w) |W|^2 + g^2 \varphi^{2}|W|^2 \nonumber \\
             &   & - 2\lambda \varphi^{2}_{o}\varphi^2 + 2g^{2}|W|^4 + \lambda
              (\varphi^4 + \varphi^{4}_{o}) \label{3.1}
\eea
\noindent
We are here ignoring the ``kinetic'' terms $(\partial_i \varphi)^2$
and $|D_i W_j - D_j W_i|^2$ and the spatial variation in $f_{12}$
in order to present some heuristic
arguments for the symmetry restoration for large external fields.
A further approximation is made by the ansatz $W_{\m}=(W,iW,0,0)$.
In the former section we saw that this was a good approximation
close to the the lower critical field strength $H^{(1)}_{crit}$.
It is not clear that this remains true away from $H^{(1)}_{crit}$,
but it can be shown to be the case \cite{6}.

If $ef_{12}$ is less than $m^2_w = \frac{1}{2}\varphi_0^2 g^2$
a minimum of $V(\varphi,W)$ is given by $\varphi = \varphi_0,~W=0$
and we have no $W$-condensate.
If $f_{12}$ is above $H^{(1)}_{crit} = \frac{1}{2}g^2\varphi^2_0$
we will get a $W$-condensate and minimizing with respect to $W$
gives
\beq
g^{2}|W|^{2}_{max} = ef_{12} - \frac{g^{2}\varphi^{2}}{2} \label{3.2}
\eeq
\noindent
As $|W|^2$ increases with $ef_{12}$ the expectation value $\varphi^2$
will decrease from $\varphi^2_0$ because the term
$g^2\varphi^2|W|^2$ will contract the Higgs term $-2\lambda\varphi^2_0
\varphi^2$.
One finds
\bea
\varphi^2_{min} & = & \varphi^2_0~ \frac{m^{2}_h - ef_{12}}{m_{h}^2 - m^2_w}
                                   \label{3.3} \\
m^2_h & \equiv &  4\lambda \varphi^2_0 ~~~~,~~~~
m^2_w \equiv  \frac{1}{2}g^2 \varphi^2_0    \label{3.4}
\eea
The Higgs- and the $W$-mass in the ordinary vacuum are denoted
$m_h$ and $m_w$ and we see that the expectation value $<\varphi^2>$
will approach zero as the average electromagnetic field strength is
larger than the Higgs mass {\it provided} the Higgs mass is larger
than the $W$ mass.
These heuristic arguments suggest that a $W$-condensate should
exist for
\beq
H^{(1)}_{crit} <{f}_{12} < H^{(2)}_{crit} \label{3.5}
\eeq
\noindent
where
\beq
eH^{(1)}_{crit} = m^2_w~~~~,~~~~eH^{(2)}_{crit} = m^2_h \label{3.6}
\eeq
and above $H^{(2)}_{crit}$ the $SU(2)_w \times U_y(1)$ symmetry
should be restored although the present simplified arguments do not
suggest what happens to the $W$-condensate above $H^{(2)}_{crit}$.

For $m_h < m_w$ we get by a minimalization of \rf{3.1} that $\varphi = 0$ as
soon as $f_{12} > H^{(1)}_{crit}$. We see that $m^{-1}_H$ and $m^{-1}_w$ are
very much like the coherence length and the penetration length of an
ordinary superconductor where $m_h = m_w$ is the borderline between
type I ad type II superconductors.
However, the analogy should not be pushed too far. Contrary to the $\psi$
condensate in a superconductor the $\varphi$ field does not couple
directly to electromagnetism and it is only through the formation of a
$W$ condensate that the symmetry restoration takes place in the electroweak
theory.

\vspace{12pt}
\noindent
At this point one can indeed ask what happens to the $W$ condensate
above $H^{(2))}_{crit}$. According to our approximate formulas above
the strength of the condensate will continue to grow all the way up to
$H^{(2))}_{crit}$. The answer requires a more detailed analysis and we have
to refer to \cite{6} for a review. It is however easy to explain
what happens: In the full theory we will not only have a condensate of
$W$'s but also of $Z$'s. In case we return to variables which are
natural in the unbroken $SU(2)_w \times U(1)_y$ theory and introduce
the vector field
$B_\mu$ for the group $U_y (1)$ and the non-abelian vector fields
$A^a_\mu$ for the group $SU(2)_W$ we have
\bea
B_\mu   &=& A_\mu \cos \theta - Z_\mu \sin \theta \label{4.26} \\
A^3_\mu &=& A_\mu \sin \theta + Z_\mu \cos \theta \label{4.27}\\
A^1_\mu &=& \frac{1}{\sqrt{2}}(W_\mu + W^\dagger_\mu) ~,~ A^2_\mu =
\frac{i}{\sqrt{2}}(W_\mu - W^\dagger_\mu)\label{4.28}
\eea
and
\bea
B_{\mu \nu}& =& \partial_\mu B_\nu - \partial_\nu B_\mu \label{4.29}\\
F^a_{\mu \nu}& =& \partial_\mu A^a_\nu - \partial_\nu A^a_\mu - g
\epsilon^{abc} A^a_\mu A^c_\nu \label{4.30}
\eea
It can now be shown (\cite{6}) that {\it when we approach $H^{(2)}_{crit}$
from below and $\vf \to 0$ the condensate solution fulfill $F_{ij}^a \to 0$.
The only non-vanishing field-strength component  at the transition
is thus the $B_{\m\n}$ associated with $U(1)_y$.}

A lesson learn from the above analysis in terms of the unbroken
variables is the importance of gauge invariance. As the condensate
solution rotates from being  mainly excitations around $U_{em} (1)$
to being mainly excitations around $U_Y (1)$ the magnitude of $W$ and $Z$
continue to increase. However, as we get close to $H^{(2)}_{crit}$
a larger and larger fraction of this increase is a pure gauge artifact
and eventually the whole "condensate" is a gauge artifact. This is why
it is  dangerous to use heuristic estimates like the ones
above in the case of gauge theories. The prediction of a symmetry
restoration turned out to be correct (\cite{6}),
but one would have been misled
with respect to the importance of the $W$ condensate above the transition
had it not been for the fact that one can analyze the situation in
full detail (\cite{6}).

\vspace{.7cm}

{\Large{\bf 5. Cosmological considerations}}
\vspace{.5cm}

The field necessary for generating the instability is given by \rf{2.12}.
Introducing ordinary egs units one finds
\beq
H^{(1)}_{crit} \approx 10^{24}\; {\rm Gauss}. \label{5.1}
\eeq
which is very large and is hard to achieve in earthbound laboratories.
There exists a slight possibility for observing the phenomenon in
high-energy (TeV) proton-proton collisions \cite{ao}, but the time needed
for the collision  may be too short to induce the instability.

Another possibility exists in the evolution of the universe. Galactic
magnetic fields of the order $10^{-6}$ Gauss have been observed in a
number of galaxies. It is usually assumed that such fields are generated
by a galactic dynamo mechanism, which amplifies a weak but coherent
primordial field of the order $10^{-21}$ Gauss or larger, on a comoving
scale of 100 kpc (see i.e. \cite{rss}). This seed field appears to be
extremely small, and does not look as a candidate for the critical field
\rf{5.1}. However, one should bear in mind that the present universe
has evolved from a much smaller universe. Since in general relativity the
magnetic flux is conserved, it follows that the magnetic field behaves
like $1/a(t)^2$, where $a(t)$ is the cosmological scale factor.
Hence, a weak field today can emerge from a strong field in the past.

It is natural, as a first approximation to assume that the field is random.
Let us consider the flux $\Phi$ through an area $L\times L$,
consisting of $L^2$ smaller areas. The flux makes a random walk, so
its size is of the order  $\sqrt{L^2}=L$ in the $L\times L$ area. Hence
the magntic field varies like $1/L$. Thus the primordial field is expected
to have the form:
\beq
H \approx \frac{M^2}{ a(t)^2 L}, \label{5.3}
\eeq
where, for dimensional reasons, $M$ is a quantity with dimensions of a mass.
Now, magnetic fields start to exist at the electroweak phase transition
$T \sim m_w$, where the natural scale is $m_w$. We can then write
$M= m_w \sqrt{c}$, where $c$ is some constant,
\beq
H\approx \frac{c\; m^2_w}{  a(t)^2\; L}. \label{5.4}
\eeq

We can now ask how large $c$ should be in order to produce a present day
primordial field of the order $10^{-21}$ Gauss at a scale of 100 kpc.
Using that $a(t)$ is proportional to $1/T$ one obtains \cite{olesen}
\beq
c \sim 10^9. \label{5.5}
\eeq
This means that at the electroweak phase transition
\beq
H_{ew} \sim 10^{33} {\rm Gauss}, \label{5.6}
\eeq
which exceeds the critical field \rf{5.1}
by many orders of magnitude. The estimate \rf{5.6} does not take into
account that there could be some unknown large scale amplification.
Such a mechanism is not assumed in the standard dynamo explanation of the
galactic magnetic fields.

By taking into account the appropriate cosmological boundary condition
it has been demonstrated by Vachaspati \cite{vac} that at the electroweak
phase transition fields of order $10^{23}-10^{24}$ Gauss are naturally
obtained. How to get the much larger field \rf{5.6} apparently needed
to explain the dynmo effect is at presently unknown, but it could be that
inflation may be a candidate for producing \rf{5.6}.

A consequence of a field as  large as \rf{5.6} is that the electroweak
phase transition is very slow. The Higgs field may not have reached
its present value before around the $QCD$ phase transition, simply
because the $W$-condensate generated by \rf{5.6} counteracts the breaking
of $SU(2)_w \times U(1))_y$ for a long time.

\end{document}